# Experimental research of longitudinal ultrasound absorption in intermediate state of high pure type I superconductor

A. G. Shepelev, O. P. Ledenyov and G. D. Filimonov

*National Scientific Centre Kharkov Institute of Physics and Technology, Academicheskaya 1, Kharkov 61108, Ukraine.*

The experimental research of the longitudinal ultrasound absorption in an intermediate state of the high pure *Ga* single crystal at the frequencies of *30 – 130 MHz* at the temperatures of *0.4 – 0.5 K*, using the impulse method, is completed. The new effects in the absorption of ultrasound in an intermediate state of the high pure *type I* superconductor are discovered. The giant oscillations in the dependence of the ultrasound absorption on the magnetic field *Γ(H)* at the magnitudes below the critical magnetic field, $H \lesssim H_c$, in an intermediate state of the high pure *Ga* single crystal are experimentally observed The maximum of the monotonic part of the ultrasound absorption on the magnetic field *Γ(H)* is also obtained. The additional experimental results in the dependence of the ultrasound absorption on the frequency of longitudinal ultrasonic wave *f* and on the orientation of external magnetic field *Γ(φ)* are reported. In the case of the high frequencies ultrasonic signal, the different behaviour of the monotonous part of the dependence of the ultrasound absorption on the magnetic field *Γ(H)* in an intermediate state, comparing to the theoretical prediction [10], is found. The anomalous distinction in the dependence of the ultrasound absorption on the orientation of magnetic field *Γ(φ)* in an intermediate state of *Ga* in comparison with the dependence of the ultrasound absorption *Γ(φ)* in a normal state of *Ga* at the same temperature and at the magnitude of magnetic field equal to the critical magnetic field, $H = H_c$, is observed. The possible theoretical mechanisms to explain the nature of big oscillations in the dependence *Γ(H)* in an intermediate state of the high pure *type I* superconductor are proposed.

## Introduction

The research results of the ultrasound absorption in high pure superconductors are successfully used to research the energy gap in the energy spectrum of superconductors and its anisotropy (see the review [1]). At the same time, the research of the intermediate state (*IS*) in the high pure type *I* superconductors, using the ultrasound, did not lead to any significant results (see the review [2]).

As it is known [3-5], the *type I* superconductors of final dimensions transit to an intermediate state, dividing on a big number of altering superconducting and normal layers of the at an action by the external magnetic field **H** with the magnitude of $(1-n)H_c < H < H_c$ (*n* is the demagnetization factor of a sample). The field **H** is equal to the critical magnetic field value $H_c$ in the normal phase layers, and it is directed along the normal phase layers; whereas there is no any magnetic field in the diamagnetic superconducting phase layers.

All the information on the *IS* is obtained by the precise observation and measurement methods, observing the intermediate phase layers appearance on the surface of researched samples (see, for example, the review [6]), because a big number of the physical properties of superconductor in an intermediate state (the electrical properties such as electrical conductivity; magnetic properties; thermodynamic properties; etc.) practically almost do not depend on the spatial period of the *IS* structure, but on the phases concentration only. In view of the fact that the *Superconductor-Normal Metal-Superconductor* (*S-N-S*) layers are deformed (the bending, branching [3-5], twisting [7]) near the surface of a sample, the advantages by the ultrasonic research of an intermediate state in the interior of superconductor are obvious due to the penetration ability by the ultrasound.

Until the recent time, the lack of significant progress in this field of research can be explained as a result of the existing difficulties toward the *IS* creation [8] and the completion of ultrasonic research toward the periodic structure of an *IS* in the interior of massive bulk superconductor, and in addition, the corresponding theoretical representations were only developed in 1966-1967. It is necessary to note that the theory is only created for the model of superconductor with the simple single-connected *Fermi* surface at this time.

In the research works by *Andreev, Bruk* [9] and *Andreev* [10], the new mechanisms of ultrasound absorption in an *IS* of superconductor were introduced and a number of new phenomena were predicted. In the first mentioned research [9], the contribution into the low frequency ultrasound absorption, connected with the vibration of inter-phase boundaries in the relatively dirty superconductor at the given relation between the electron mean free path $l$ and the thickness of normal phase layer $a$: $l<<a$, is researched. In the second research work [10], the influence by the specific *Andreev* reflections [11] of electronic excitations ("electron" and "hole") at the inter-phase boundaries in an intermediate state of high pure superconductor at $l>>a$ in the range of high frequencies of ultrasonic signal on the high frequency ultrasound absorption is investigated. The matter is that, at $T<T_c$, the electron excitations of normal phase have the energy, which is small, comparing to the energy gap in the superconducting phase, and are fully reflected on the inter-phase boundary. The specifics of these reflections is concluded in the fact that the value of electronic excitation's impulse is preserved, but the sign of electronic excitation's velocity vector, mass, and charge are changed on the opposite one.

In view of significant experimental difficulties, there were no data about the direct detection and discovery of theoretically predicted effects [9, 10].

The most interesting from them – the oscillations of longitudinal ultrasound absorption in an *IS* of superconductor at $k \perp H$, the period of which slowly changes with the variation of normal phase layer thickness $a(H)$ at the change of magnitude of external magnetic field $H$ ($k$ is the wave vector of ultrasonic wave). During this process, in agreement with the theory [10], the kinetic equation for the electrons in this layered medium at the presence of both the ultrasonic oscillations and the magnetic field coincides with the kinetic equation for the electrons in the same conditions in the normal metal. At $ka \sim 1$, it is possible to conduct some analogy between the predicted oscillations of ultrasound absorption in an *IS* of superconductor and the usual magneto-acoustic oscillations in a normal metal. As it is known, in the last phenomenon, at a change of the extreme diameter of electron's orbit at the magnetic field $D_{ext}(H)$, the conditions of effective interaction of electron with the ultrasonic oscillations with the wavelength $\lambda = const$ appear periodically. In the new phenomenon, the effective interaction of ultrasonic waves with the electron system is periodically dependent on the normal phase layer thickness $a(H)$, while $D_{ext}(H_c) = const$. At this, the value of $2a(H)$ plays a role of the parameter, which can be compared with the wavelength $\lambda$ (in the previous effect). The new phenomenon can not exist without the presence of periodic *S-N-S* layered structure in an *IS* and without the fulfillment of a number of strong inequalities between the electron mean free path $l$, normal phase layer thickness $a$, and extreme diameter of electron's orbit in the normal phase layer $D$: $l>>D_{ext}>a$. The presence of usual magneto-acoustic oscillations in a normal state of superconductor at $T>T_c$ and $H<<H_c$ serves as main criteria of most hard condition fulfillment: $l>>D$, when the magnetic field $H$ has its defining influence on the electrons dynamics in the normal phase layers.

The three experimental attempts to detect the oscillations of ultrasound absorption in an *IS* of high pure *Tin* are well known. In the research work by *Bezugluy et al.* [12], the dependence of the longitudinal ultrasound absorption with the frequency of *1–110 MHz* on the magnitude of external magnetic field $H$ with the constant orientation at the constant temperature of *1.5K* (at $k \perp H$) was researched. In our view, the phenomenon was not detected, because the condition: $l>>D_{ext}$ was not fulfilled. It is necessary to note that, in the beginning, the researchers [12] explained the failure of their experiments, because the extreme diameter of electron orbit was bigger than the thickness of normal phase layer: $D_{ext}>a$. However, this conclusion is founded on the misconception, because the electron experiences the *Andreev* reflections on the inter-phase boundaries at the fulfillment of the condition: $D_{ext}>a$, thus one of the necessary conditions of presence of the oscillations of ultrasound absorption in an *IS* of superconductor is satisfied.

In the next research work by *Bezuglyi et al.* [13], the idea that the period of oscillations of ultrasound absorption, which is to be discovered, must be bigger than all the range of change of magnetic field $H$ was provided as an explanation of negative obtained result in [12]. However, the analysis of this physical problem with the detailed consideration of the data on the *Fermi* surface of the *Tin* [14-17] show that the periodicity of the phenomenon, in the case of its existence, will only take some intervals in the full range of change of the magnetic field $H$.

In [13], the dependence of the longitudinal ultrasound absorption with the frequency of *19.5 MHz* (at $k \perp H$) on the magnitude of critical magnetic field $H_c = H_c(T)$ in the normal phase layers in an intermediate state of the *Tin* was researched; it makes sense to note that the researchers tried to keep the normal phase layer thickness $a$ at the constant level during the changing temperature in the range of the temperatures from *2.1K* up to *3.37K*. Of course, the shown weak non-monotonicities of ultrasound absorption [13] can not be considered as the confirmation of theoretically predicted phenomenon. In view of the fact that, in [12, 13], the special measures to create the fine periodic geometric structure in an IS of superconductor were not undertaken, hence the authors fairly commented on a possible existence of the normal phase layers with the different thickness in a researched sample (p. 250 in [13]). Naturally, in the experiment [13], the usual magneto-acoustic oscillations were registered in the normal phase layers with the thickness of $a \gtrsim D_{ext}$; here, it is necessary to comment that the condition of existence of the magneto-acoustic oscillations: $l \leq D_{ext}$ [18] can be more easily realized in the normal metal. Therefore, at the completion of experimental measurements as in the considered experiment in [13], it is necessary to be confident that the observed "non-monotonicities" are not related to the usual magneto-acoustic oscillations, which occur at the change of the $H_c(T)$ in the normal phase layers of a superconducting sample. Let us note that the first research work [12] was free of this significant misassumption, because its experimental conditions – the low enough temperature and high enough frequency of ultrasonic signal – made it possible to come close to the following condition fulfillment: $l >> D_{ext}$, resulting in the high sensitivity of measurements.

In the experimental research work by the *Canadian* researchers [19], which was conducted at the constant temperature in the frequencies range of longitudinal ultrasonic signal from *11 MHz* up to *110 MHz* (at $k \perp H$), the theoretically predicted phenomenon was not discovered despite of the special measures undertaken to create the fine periodic structure of an *IS* of the *Tin*. This is a one more confirmation of the highlighted fact that the theoretically predicted phenomenon can not originate in the *Tin*, if the condition: $l >> D_{ext}$ is not fulfilled.

The high pure *Gallium* single crystal is one of a limited number of the high pure *type I* superconductors, which satisfies all the requirements toward the existence of the oscillations of ultrasound absorption in an intermediate state of superconductor. The usual magneto-acoustic oscillations can be observed in a normal state of the high pure *Ga* single crystal, starting at the external magnetic fields *H* of several *Oersted* [20, 21]. For the first time, the giant oscillations of longitudinal ultrasound absorption in an *IS* of high pure *type I* superconductor were experimentally discovered by *Shepelev, Ledenyov, Filimonov* in [22]. These oscillations are well described, using the theory by *Andreev* [10]. At the same time, it was found that all the experimental results can not be fully explained. Therefore, the further experimental and theoretical research works have to be continued in this field of research.

**Experimental research measurements**

In the present research[1], the additional information on the dependence of the ultrasound absorption on the frequency of longitudinal ultrasonic signal *f* and on the orientation of magnetic field *H* in an *IS* of high pure *Ga* single crystal is reported[2].

The experimental measurements on the precise characterization of physical properties of high pure *Ga* single crystal[3] were conducted in the cryostat with the liquid $^3He$ [23] at the frequencies of $30-130MHz$ at the temperatures of $0.4-0.5K$, using the impulse method. The electron mean free path in the high pure *Ga* single crystal is $l \sim 1 cm$. The axis of cylindrical sample coincided with the crystallographic axis *b* and wave vector of longitudinal ultrasound *k*. In the process of measurements, the sample with the diameter of *7 mm* and length of *21 mm* was in a direct contact with the liquid $^3He$, and the magnetic field of the *Earth* was precisely compensated by the two pairs of the *Helmholtz* coils.

The *IS* of superconductor was created in a sample, using the homogenous transverse external magnetic field *H* (the homogeneity of the magnetic field *H* was better than $10^{-3}$), generated by the one additional pair of the *Helmholtz coils*, herewith, the magnetic field vector *H* was able to rotate in the plane of the *Ga* crystallographic axes *a* and *c* with the velocity of *1 rev/min*.

As it was explained in [22, 25, 26], the rotation of the magnetic field *H* stipulates the generation of the fine intermediate state structure, which is close to the equilibrium structure. Indeed, the dependence of the ultrasound absorption in an *IS* of the pure *Ga* at the ultrasonic signal frequency of *130 MHz*, which is obtained without the rotation of magnetic field *H* (Fig. 1) differs from the dependence of the ultrasound absorption in an *IS*, obtained at the same frequency, ultrasound propagation direction, temperature and magnetic field orientation $\angle H, c = 22°$, but with the rotation of the magnetic field *H* between the neighboring points (in Fig. 2) [22]. Let us note that the oscillations of ultrasound absorption in an *IS*, reported in the present research paper (see Fig. 1), are observed in the region with the big concentrations of normal phase at the conditions, which are close to the equilibrium conditions, as described in Fig. 1 in [22]. In the both considered experimental cases, the monotonous part of ultrasound absorption dependence has a type of dependence, which can not be explained presently.

Going from the existing theoretical representations, the monotonous part of ultrasound absorption dependence can be approximated by the linear type of dependence in the form of the direct line in the limit of $ka >> 1$, which makes the connection between the beginning point and the end point of an *IS* of superconductor (see Fig. 1). However, in the experiment, the magnitude of ultrasound absorption in an *IS* at the given ultrasonic signal frequency and magnetic field orientation is significantly bigger than the theoretical limit, and it is even bigger than the full electron absorption in the sample in the normal state at $H = H_c$.

In Fig. 3, the dependence of the ultrasound absorption on the magnetic field $\Gamma(H)$ at the same frequency of *130MHz* and same ultrasound propagation direction, obtained at the temperature of *0.4K*, but at the changed orientation of magnetic field $\angle H, c = 66°$ is shown. Naturally, the new orientation of the magnetic field *H* could have an impact on the magnitude of ultrasound absorption in a normal state of superconductor $\Gamma_0^N(H_c)$, however, in agreement with the theory [10], the shape of the monotonous part of the dependence of the ultrasound absorption on the magnetic field $\Gamma(H)$ in an *IS* of superconductor, must not change.

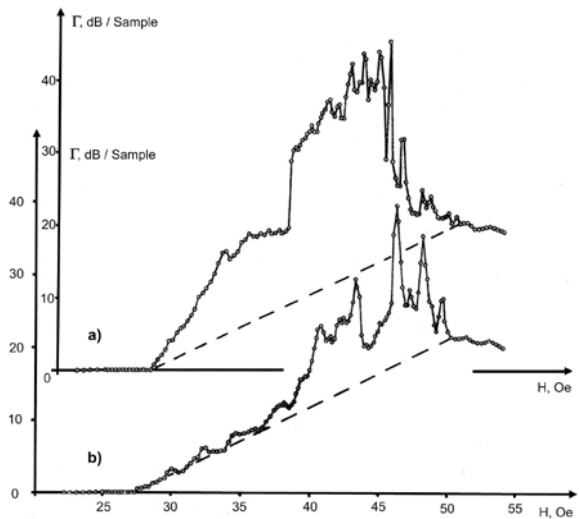

*Fig. 1. Dependence of longitudinal ultrasound absorption on magnetic field Γ(H) in intermediate state of high pure Ga single crystal at frequency of 130 MHz at temperature T = 0.4 K, **k** // **b**, ∠**H**, **c** = 22°:*
*a) transition from superconducting state to normal metal state;*
*b) transition from normal metal state to superconducting state.*
*Interval of every measurement is 2 hours; rotation of magnetic field **H** between the points was not conducted. Dashed lines represent monotonous part in dependence Γ(H) at ka>>1 in agreement with theory [10].*

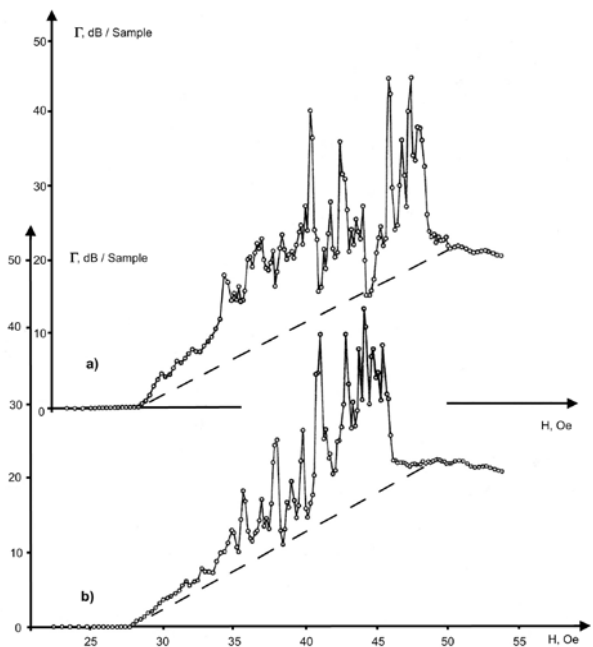

*Fig. 2. Dependence of longitudinal ultrasound absorption on magnetic field Γ(H) in IS of high pure Ga single crystal at frequency of 130 MHz at temperature T = 0.4 K, **k** // **b**, ∠**H**, **c** = 22°:*
*a) transition from superconducting state to normal metal state;*
*b) transition from normal metal state to superconducting state. Interval of every measurement is above 5 hours; two rotation of magnetic field **H** between the nearest points were made.*
*Interval of every measurement is above 5 hours; two rotation of magnetic field **H** between the nearest points were made.*
*Dashed lines represent monotonous part in dependence Γ(H) at ka>>1 in agreement with theory [10].*

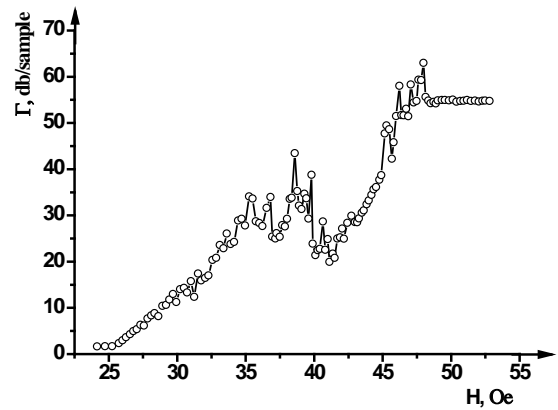

*Fig. 3. Dependence of longitudinal ultrasound absorption on magnetic field Γ(H) in intermediate state of high pure Gallium single crystal at transition from superconducting state to normal state at frequency of 130 MHz at temperature T=0.4K, **k** // **b**, ∠**H**, **c** =66°: Interval of every measurement is around 5 hours; one full rotation of magnetic field **H** between the nearest points with the velocity of 1 rev/min was made.*

The experimentally found distinction in the monotonous part of the dependence of the ultrasound absorption on the magnetic field $\Gamma(H)$ is probably connected with the multi-connectivity of the *Fermi* surface of researched superconductor, which was not taken to the consideration in the theory [10], because the only strongly changed parameter at the altered orientation of magnetic field *H* is the extreme diameter of electrons orbit $D_{ext}$ on the *Fermi* surface in the normal phase layers in an *IS* of superconductor. However, it is too early to make any final conclusions on the real reasons of observed physical behaviour of the monotonous part of the dependence of the ultrasound absorption on the magnetic field $\Gamma(H)$ in an *IS* of the high pure *Ga* single crystal. It looks like it is necessary to take into the consideration the extreme conditions of our experiment: $l>>D_{ext}$, $l>>a$ and $ka>>1$. Moreover, it is not clear: what is the contribution by the physical mechanism of the inter-phase boundaries vibration [9] in the magnitude of the high frequency ultrasound absorption, because the overall impact of this mechanism is considered in application to the different conditions: the frequencies of $10^6$–$10^7$ Hz and $l<<a$.

We also conducted the experiments with the high pure *type I* superconductor at the frequency of *30 MHz* the temperature of *0.5 K* and the orientation of magnetic field ∠*H*, *c* = 22°. In Fig. 4, the automatic recording of the amplitude of ultrasonic signal, which propagated through the sample, shows that there are the big oscillations of ultrasound absorption at the big concentrations of normal phase in an *IS* of the high pure *Ga* single crystal at the four times lower frequency than the above mentioned frequency [22] and at the high velocity of change of the magnitude of magnetic field *1.7 Oe / min.*

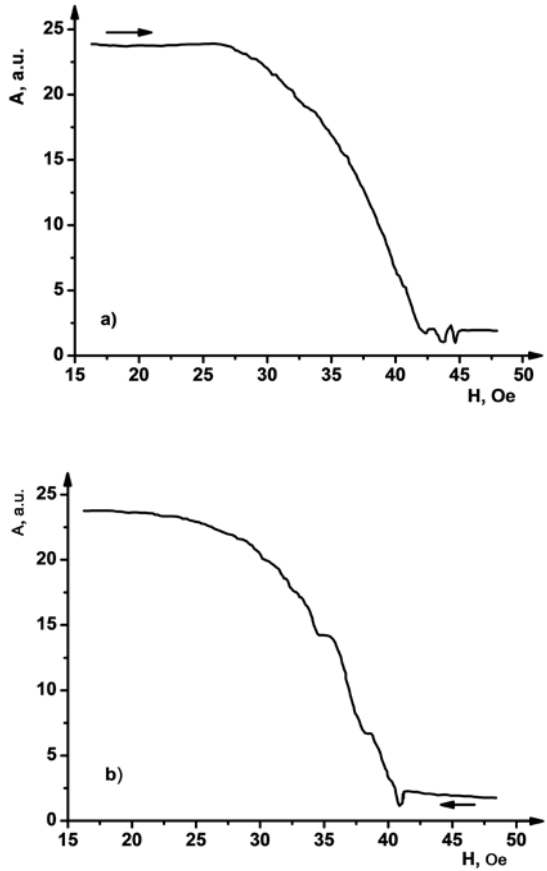

*Fig. 4. Dependence of amplitude of ultrasonic signal on magnetic field Γ(H) in IS of high pure Gallium single crystal at frequency of 30MHz at temperature T=0.5K, **k** // **b**, ∠**H**, **c** = 22°.*
*Velocity of change of magnetic field H is 1.7 Oe / min.*
*a) transition from superconducting state to normal metal state;*
*b) transition from normal metal state to superconducting state.*

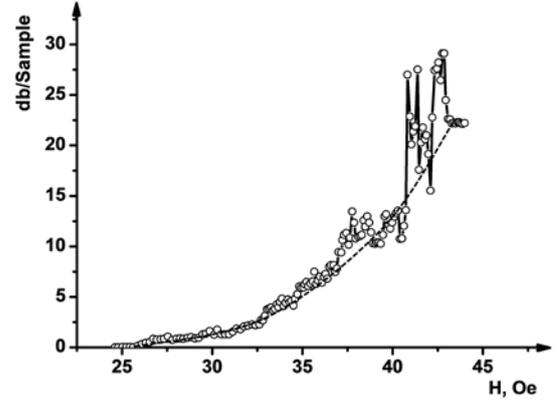

*Fig. 5. Dependence of longitudinal ultrasound absorption on magnetic field Γ(H) in IS of high pure Ga single crystal at transition from superconducting state to normal state at frequency of 30MHz at temperature T = 0.5 K, **k** // **b**, ∠**H**, **c** = 22°. Interval of every measurement is around 9 hours; two full rotations of magnetic field **H** between the nearest points with the velocity of 1 rev / min were made.*
*Dashed line represents monotonous part in dependence Γ(H) in agreement with theory [10].*

Of course, these oscillations of ultrasound absorption are more clearly shown in the dependencies of the ultrasound absorption on the magnitude of magnetic field *Γ(H)*, which are created, using the measurements at the selected points at the frequency of *30 MHz* at the temperature of *0.5 K*, as shown in Figs. 5 and 6.

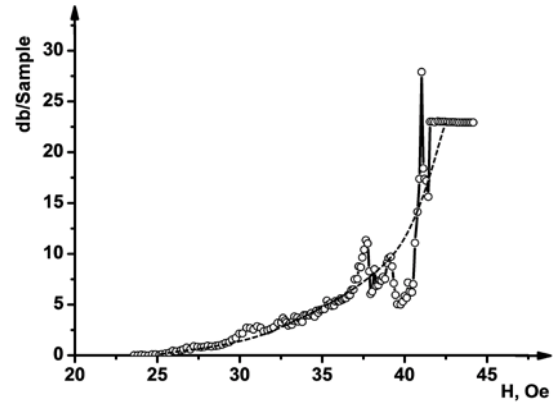

*Fig. 6. Dependence of longitudinal ultrasound absorption on magnetic field Γ(H) in IS of high pure Ga single crystal at transition from normal state to superconducting state at frequency of 30 MHz at temperature T = 0.5 K, **k** // **b**, ∠**H**, **c** = 22°. Interval of every measurement is around 9 hours; two full rotations of magnetic field **H** between the nearest points with the velocity of 1 rev / min were made. Dashed line represents monotonous part in dependence Γ(H) in agreement with theory [10].*

The time period of uninterrupted measurements was increased up to *9* hours, giving an opportunity to get the more precise measured data. The two full rotations of the magnetic field vector *H* with the velocity of *1 rev/min* were conducted in every point of measurement, and the increase of magnetic field between the points was $\Delta H = 0.002 H_c$. The velocity of change of magnetic field between the neighboring points was *0.7 Oe/min*.

In Figs. 5 and 6, the monotonous parts of curves are well described by the theory, thus it is possible to find the thickness of the normal phase layer in the interior of superconductor, using a different method; and then to compare it with the thickness of normal phase layer, calculated with the use of the measured period of the oscillations of ultrasound absorption in an *IS* of the high pure *type I* superconductor [22]. In the theory [10], the monotonous part of ultrasound absorption in an *IS* can be described by the following equation

$$\Gamma = \eta \, \Gamma_0^N (H_C) \Phi \left( \frac{ka}{\pi} \right)$$

where $\eta$ is the concentration of the normal phase, $\Gamma_0^N(H_c)$ is the magnitude of the monotonous part of ultrasound absorption in the normal metal state in the magnetic field $H_c$, $\Phi(ka/\pi)$ is the function with the numerical values given in [10]. Thus, having obtained the variables: $\Gamma$, $\Gamma_0^N(H_c)$ from the experiment, it is possible to get the function $\Phi(ka/\pi)$; and knowing the $k$, it is possible to find the thickness of normal phase layer $a$. The normal metal phase layer thickness in an *IS* of high pure *Ga* single crystal in the case of big concentrations of normal metal phase, measured at the propagation of ultrasonic signal with the frequency of *30 MHz*, is $a = 10^{-2} cm$, and it is equal to the value of normal metal phase layer thickness, which was calculated in [22], using the period of oscillations of ultrasound absorption at the frequency of *130 MHz*.

It is interesting to note that the amplitude of the oscillations of ultrasound absorption at the frequency of *30 MHz* in an *IS* of the high pure *Ga* single crystal at the big concentrations of normal phase is close to the value of change of the magnitude of longitudinal ultrasound absorption at the same frequency in the "rotation diagram" $\Gamma(\varphi)$ (see Fig. 3 in [27]). In Figs. 5 and 6, the experimental curves are obtained in an *IS* of the *Ga* single crystal at the certain orientation of the external magnetic field *H* in relation to the crystallographic axes of the crystal, by the way of the change of the normal phase layer thickness $a$ with the change of the magnitude of the magnetic field *H*; while the "rotation diagram" $\Gamma(\varphi)$ [27] was obtained in an *IS* by the way of change of the orientation of the external magnetic field *H* in relation to the crystallographic axes, but without the change of the magnitude of the external magnetic field *H*, at the constant value of the normal phase layer thickness $a$. The magnitude of the oscillations of ultrasound absorption in the first case; and the range of change of the magnitude of longitudinal ultrasound absorption in the second case, are comparable with the magnitude of the full electron ultrasound absorption, and they can not be explained somehow differently rather than as in agreement with the *Andreev* theory [10]. Indeed, in the first case of the oscillations of ultrasound absorption: $\Delta \Gamma(H) \sim 10 dB/Sample$, in the second case: $\Delta \Gamma(\varphi) \sim 10 dB/Sample$, while the magnitude of the electron ultrasound absorption in a normal state of superconductor at the external magnetic field $H = H_c$ at the given conditions of experiment is $\Gamma_0^N(H_c) = 23 dB/Sample$. Moreover, the full electron ultrasound absorption in the high pure *Ga* single crystal at this frequency at the selected propagation direction of ultrasonic signal, which was obtained due to the measurements of the temperature dependence of ultrasound absorption at the temperatures below the critical temperature $T < T_c$ without the application of the external magnetic field *H*, is close to the following magnitude: $\Gamma^N \approx 24 dB/Sample$.

## Conclusion

In the cycle of research works, presented in [22, 27] and in this research paper, for the first time, the new effects in absorption of ultrasound in an intermediate state of high pure *type I* superconductor at $k \perp H$ are discovered:

1. In agreement with the theory [10], the giant oscillations of ultrasound absorption in an *IS* of high pure *Ga* single crystal at the frequencies of *30 - 130 MHz* at the temperatures of *0.4 – 0.5 K* at the different orientations of the magnetic field *H* are detected experimentally [22]. These oscillations of ultrasound absorption are originated, because of the presence of the periodic *S-N-S* structure in an *IS* of the high pure *type I* superconductor. The appearance of the *S-N-S* structure in an intermediate state of the *type I* superconductor was predicted by *Landau* [3]. The thickness of normal phase layer at the big enough concentrations of normal phase in the interior of superconductor: $a = 10^{-2} cm$, was found, using the measured period of oscillations of high frequencies ultrasound absorption.

2. The anomalous distinction in the dependence of the ultrasound absorption on the orientation of magnetic field $\Gamma(\varphi)$ in an intermediate state of superconductor in comparison with the dependence of the ultrasound absorption on the orientation of magnetic field $\Gamma(\varphi)$ in a normal state of superconductor at the same temperature and at the magnitude of magnetic field equal to the critical magnetic field, $H = H_c$, is found [27]. The origination of this effect is stipulated by the existing multi-connectivity of the *Fermi* surface in the *Ga* single crystal, and it was not considered theoretically. Let us note that this effect can not appear in the metals with the isotropic *Fermi* surface.

3. During the research on the high frequencies ultrasound absorption, the different physical behaviour of the monotonous part of the dependence of the ultrasound absorption on the magnetic field $\Gamma(H)$ in an *IS* of *Ga* single crystal, comparing to the theoretical prediction [10], is found. The precise characterization of the monotonous part of the dependence of the ultrasound absorption on the magnetic field $\Gamma(H)$ is completed at the two selected frequencies of ultrasonic signal $f_1$ and $f_2$ at the two selected orientations of magnetic field $H_1$ and $H_2$. We assume that it is necessary to take to the consideration both the complexity of the *Fermi* surface in the researched high pure *Ga* single crystal and the extreme conditions of our experiment: the high frequencies of ultrasonic signal; the fact that the electron mean free path is much more bigger than the extreme diameter of electron orbit: $l >> D_{ext}$, the thickness of normal phase layer *a*, and the strong inequality expression: $ka >> 1$.

4. The research on the accurate characterization of the ultrasound absorption in the high pure *Ga* single crystal at the low frequencies of ultrasonic signal shows that the monotonous part of the dependence of the ultrasound absorption on the magnetic field *Γ(H)* in an *IS* is in good agreement with the theoretical prediction. This circumstance allowed us to use a different method to find the normal phase layer thickness rather than the method in [22]. It makes sense to emphasis that the normal phase layers thicknesses in the interior of superconductor in an *IS*, which were calculated for the case of the big concentration of normal phase at the different frequencies of ultrasonic signal and by the different methods, have the same value: $a = 10^{-2} cm$.

The obtained experimental research results on the oscillatory ultrasound absorption in an *IS* of the high pure *type I* superconductor are different from the early data by other researchers (see the review [2]). The discovered new physical effects will be comprehensively investigated in our forthcoming experimental and theoretical researches.

1) The innovative experimental research results were reported at the *12 International Conference on the Low Temperature Physics and Techniques* in Hungary in 1973.
2) A. D. Stryuk, who was a student at the *Karazin Kharkov National University*, took an active part in the experimental measurements.
3) The high pure *Ga* single crystal was synthesized with the use of the *Chokhralsky* method at the *Giredmet Experimental Plant* in Russia.

This research paper was published in the *Problems of Atomic Science and Technology* in *Russian* in 1973 [28].

*E-mails: ledenyov@kipt.kharkov.ua

―――――――――